\documentstyle[aps,preprint,tighten,epsfig]{revtex}
\newcommand{\lsim}{\mathrel{\mathop{\kern 0pt \rlap
  {\raise.2ex\hbox{$<$}}}
  \lower.9ex\hbox{\kern-.190em $\sim$}}}
\newcommand{\gsim}{\mathrel{\mathop{\kern 0pt \rlap
  {\raise.2ex\hbox{$>$}}}
  \lower.9ex\hbox{\kern-.190em $\sim$}}}

\begin{document}


\title{Sensitivity plots for WIMP direct detection using the annual modulation signature}

\author{S. Cebri\'{a}n, E. Garc\'{\i}a,
         D. Gonz\'{a}lez, I. G. Irastorza, A. Morales, J. Morales, A. Ortiz de
         Sol\'{o}rzano, A. Peruzzi,
         J. Puimed\'{o}n, M. L. Sarsa,
         S. Scopel, J. A. Villar}

\address{Laboratorio de F\'{\i}sica Nuclear. Universidad de Zaragoza\\
              50009, Zaragoza, SPAIN}

\maketitle\abstract{
Annual modulation due to the Earth's motion around the Sun is a well known signature of
the expected WIMP signal induced in a solid state underground detector.
In the present letter we discuss the prospects of this technique on statistical grounds,
introducing annual-modulation sensitivity plots for the WIMP--nucleon scalar cross section for
different materials and experimental conditions. The highest sensitivity to modulation
is found in the WIMP mass interval $10 \;{\rm GeV}\lsim m_W \lsim 130\;{\rm GeV}$,
the actual upper limit depending from the choice of the astrophysical
parameters, while the lowest values of the explorable WIMP--nucleon elastic
cross-sections fall in most cases within one order of magnitude of the sensitivities of
present direct
detection WIMP searches.
}
\vspace{2 cm}


\section{Introduction}

Non barionic Cold Dark Matter is a basic component in
cosmological models of structure formation.
Taking into account the result of microlensing surveys, it
can be stated that dark baryons can account for at most only about one-third of the estimated density
of our dark halo. The best candidates to provide the rest are Weakly
Interacting Massive Particles (WIMP), among which the neutralino, provided by the Supersymmetric
extension of the Standard Model, is one of the
favourites.

It is well known that
all WIMP direct searches, which look for the WIMP elastic scattering off the nuclei
of a suitable detector, are essentially constrained by the fact that the
predicted differential rate of the signal has
a dependence with decreasing energy which is hardly distinguishable
from the background recorded in the detector. While this
fact does not prevent us in extracting upper bounds of the
WIMP--nucleus interaction cross--section for each WIMP mass, a
distinctive signature is needed to claim a positive identification of the WIMP.
In any case, the non appearance of the genuine signature
looked for
could provide a background--rejection method
useful to improve the experimental sensitivity.

The only identification signatures of the WIMP explored up to now
are provided by the
features of the Earth's motion with respect to the Dark Matter halo.
In particular
the annual modulation effect\cite{drukier} is provided by
the combination of the motion of the solar system in the Galactic rest frame
and the rotation of the earth around the Sun. Due to this effect
the incoming WIMP's velocities in the detector rest frame change continuously during
the year, having a maximum in summer and a minimum in winter.

Several experiments have already searched for this effect\cite{modza,modarg,damaxe}
one of which\cite{dama} reports a
possible positive signal.
The present situation is no doubt exciting, since
experimental sensitivities of underground detectors are entering for the first time
the supersymmetric parameter space,
and a host of new experiments will
soon start to probe it with even higher sensitivity.
In the present paper we discuss on purely statistical grounds
the perspectives of modulation searches taking into account the features of present and future
experiments. Sensitivity plots will be introduced in order to convey in a compact way all the
necessary information.

\section{\bf Extracting the modulation signal}

The procedure to extract a modulated signal with a given period and phase from a set of
measured count rates has been discussed by
several authors\cite{frees,ramachers,hasenb,dama}.

Due to the Earth's rotation around the Sun, the expected count rate
of WIMP's scatterings off the target's nucleus
changes periodically in time. The dependence may be approximated by a cosine
function with period T=1 year and phase $t_0=2^{nd}$ june:
\begin{equation}
  S=S_{0}+S_{m}\cos\omega(t_i-t_0)
\end{equation}
\noindent where $S_0$ and $S_m$ are the constant and
the modulated amplitude of the signal respectively.
The oscillating frequency is
$\omega=2\pi/T$ and the $i$ index indicates the day.
The theoretical
inputs introduced in the evaluation of the
WIMP-nucleus elastic
scattering (cross sections,
nuclear form factors, scalar, vectorial or spin--dependent nature of the coupling)
and all the parameters entering in the halo
velocity distribution (r.m.s. and escape WIMP velocity, local dark
matter density) make the evaluation of the functions $S_{0}$ and $S_{m}$ rather model
dependent \cite{jkg}. As is costumary, the amplitudes $S_0$ and $S_m$ are
expressed in terms of the WIMP mass $m_W$
and of the point--like WIMP--nucleus cross section $\sigma$.

Given a set of experimental count rates $N_{ik}$ representing the number of events
collected in the i-th day and k-th energy bin
(the formalism can be easily generalized to the case of a multiple-crystal set-up),
the mean value of $N_{ik}$ is:
\begin{equation}\label{media}
  <N_{ik}>\equiv\mu_{ik}=\left[b_{k}+S_{0,k}+S_{m,k}\cos\omega(t_i-t_0)\right]\cdot W_{ik}
\end{equation}
\noindent where the $b_{k}$ and the $S_{ik}\equiv S_{0,k}+S_{m,k}\cos\omega(t_i-t_0)$
represent the average background and the signal respectively,
in number of counts per unit of detector mass, time and interval of
collected
energy $E$ (which is related to the recoil
energy $E_R$ by the relation $E=QE_R$ where $Q$ is the quenching factor of
the detector).
$W_{ik}=M\Delta T_i\Delta E_k\epsilon_{k}$ are the
corresponding exposures, where $M$ is the mass of the
detector, $\Delta E_k$ is the amplitude of the k-th energy--bin, while
$\Delta T_i$ represents the i-th time bin (in the following we will
assume all $\Delta T_i$= 1 day).
The $\epsilon_{k}$ are efficiencies that have to be taken
into account whenever some subtraction method is used with the
data. In the following they will be neglected.
For simplicity also $t_0$ will be omitted in the following equations.

A general procedure to compare theory with experiment is to
use the maximum-likelihood method.
The combined-probability
function of all the collected $N_{ik}$, assuming that they have a
poissonian distribution with mean values
$\mu_{ik}$, is given by:
\begin{equation}
  L=\prod_{ik}
  e^{-\mu_{ik}}\frac{\mu_{ik}^{N_{ik}}}{N_{ik}!}\label{likelyhood}.
\end{equation}

Assuming that the average background rates $b_k$ and the efficiencies $\epsilon_k$
are constant in time, a possible WIMP oscillating solution can be
searched for in the data.
This implies that the time--dependent fluctuations of background and efficiencies
should stand well below the size of the searched modulation effect,
if one wishes to disregard any other less exotic explanation of a modulation of the data.
If this condition is verified, a WIMP analysis is justified,
and the most probable values of $m_W$ and $\sigma$ maximize
$L$ or, equivalently, minimize the function:
\begin{eqnarray}
  y(m_W,\sigma) &\equiv& -2 \log{L}-const  \nonumber \\
                &=& 2 \mu -2\sum_{ik}N_{ik}\log
  \left[b_{k}+S_{0,k}+S_{m,k}\cos\omega t_i\right]\label{ypiccolo}
\end{eqnarray}
\noindent where $\mu\equiv\sum_{ik}\mu_{ik}$
and all the parts not depending on $m_W$ and $\sigma$  may be
absorbed in the constant because are irrelevant for the minimization.

The function $y$ is minimized in a two--step procedure, first with
respect to the time--independent parts $f_{k}\equiv b_{k}+S_{0,k}$
and then with respect to $m_W$ and $\sigma$. In the first
minimization the $b_{k}$ are free parameters with the only
constraint to be positive. So the condition $S_{0,k}\le
f_{jk,min}$, or otherwise $S_{0,k}$=$f_{jk,min}$, is imposed.

Following a standard procedure\cite{pdg}, a region of n standard
deviations around the minimum in the plane $(\sigma,m_W)$ can be
found by imposing the condition $y(\sigma,m_W)-y_{min}\le n^2$.

\section{Data reduction}
Although the likelihood function $L$ depends on
all the collected data $N_{ik}$, only suitable
combinations of the $N_{ik}$ enter in the
determination of the parameters $\sigma$ and $m_W$. By expanding $y$ in
powers of $x_{k}\equiv S_{m,k}/f_{k}\lsim$ few \%
up to terms of the order $x_{k}^2$ the following expression can be
obtained (here and in the following $Var$ indicates the variance):
\begin{eqnarray}\label{eq:sviluppo}
 y(\sigma,m_W)&=&\sum_k
 \frac{(S_{m,k}(\sigma,m_W)-X^{\prime}_{k})^2}{Var(X^{\prime}_{k})}+{\cal
 O}(x_k^3)+F(N_{ik})\\
 X^{\prime}_k&\equiv& \frac{\sum_i N_{ik} \cos{\omega t_i}-N_{k}\beta_{k}}{W_k (\hat{\alpha}_k
 -\hat{\beta}_k^2)}\;\;\;\; Var(X_k^{\prime})\equiv
 \frac{N_k}{W_k^2(\hat{\alpha}_k-\hat{\beta}_k^2)}\label{eq:xprime}
\end{eqnarray}
\noindent where
 $\beta_k\equiv \frac{\sum_i W_{ik} cos \omega
 t_i}{W_k}$, $\hat{\alpha}_k\equiv \frac{\sum_i N_{ik} cos^2 \omega
 t_i}{N_k}$, $\hat{\beta}_k\equiv \frac{\sum_i N_{ik} cos \omega
 t_i}{N_k}$, $N_k\equiv \sum_i N_{ik}$, $W_k\equiv \sum_i W_{ik}$.
The last term of Eq.(\ref{eq:sviluppo}) may be dropped, since it
does not depend on the fitting parameters.
This implies that the likelihood function has an approximate
factorization $L\simeq F_1(N_{ik})\cdot F_2(X_k^{\prime},\sigma,m_W)$
and the information to determine $\sigma$ and $m_W$ is only
contained in the $X^{\prime}_k$ up to negligible corrections.
As expected on general grounds, Eq.(\ref{eq:sviluppo}) reduces asymptotically
to a $\chi^2$: this happens when the $X_k^{\prime}$'s have
gaussian distributions (in the examples that will be
discussed in the following we have checked that this is always verified).

The expressions (\ref{eq:xprime}) of the $X_k^{\prime}$´s are asymptotically equal to the
cosine projections introduced in ref. \cite{frees} (generalized to the case of discontinuous
data taking and unpaired days) to extract a modulated signal,
\begin{eqnarray}
 X_k &\equiv& \frac{\sum_i N_{ik} \cos{\omega t_i}-N_{k}\beta_{k}}{W_k (\alpha_k
 -\beta_k^2)}\;\; ; <X_k>=S_{m,k}\\
 Var(X_k) &\equiv&
 \frac{\sum N_{ik} cos^2 \omega t_i+\beta_{k}^2 N_k -2 \beta_k \sum N_{ik} \cos \omega
 t_i}
 {W_k^2(\alpha_k-\beta_k^2)^2}\label{eq:xfreese}\\
\end{eqnarray}
\noindent with $\alpha_k\equiv \frac{\sum_i W_{ik} cos^2 \omega
 t_i}{W_k}$. (For the sake of
comparison with Eq.(\ref{eq:xprime}) it is convenient to use the identity
$N_{k}\equiv(\sum N_{ik}
\cos^2\omega t_i+\hat{\beta}_k^2 N_{k}-2 \hat{\beta}_k \sum N_{ik} \cos \omega
t_i)/(\hat{\alpha}_k-\hat{\beta}_k^2)$ in the expression of $Var(X^{\prime}_k)$). This
implies that the
likelihood function $y$ has also the asymptotic behaviour:
\begin{equation}\label{eq:chi2}
   y(\sigma,m_W)=\chi^2(\sigma,m_W)\;\;\; ; \chi^2(\sigma,m_W)\equiv \sum_k
 \frac{(S_{m,k}(\sigma,m_W)-X_{k})^2}{Var(X_{k})}.
\end{equation}
In the case of paired days ($\beta_k=0$) and if only time
intervals
close to the maximum and the minimum of the cosine function are
considered ($\cos\omega t_i \simeq \pm 1$) the $X_k$'s reduce to
the june--december differences in the collected count rates in
each energy bin $E_k<E<E_k+\Delta E_k$.

\section{Statistical significance of the signal}

Once a minimum of the likelihood function is found, a positive result excludes absence
of modulation at some confidence level probability. This can be checked by
evaluating the quantity $\delta^2=y(\sigma=0)-y(\sigma,m_W)_{min}$ to test the
goodness of the null hypothesis.
In order to study the distribution of $\delta^2$ we make use of the asymptotic behaviour
(\ref{eq:chi2}). This implies:
\begin{equation}\label{eq:asimptotic}
  \delta^2=y(\sigma=0)-y_{min}\simeq\chi^2(\sigma=0)-\chi^2_{min}.
\end{equation}
In the case of absence of a modulation effect (i.e. time-independent Poisson-fluctuating
data) numerical simulations show that the quantity $\delta^2$ belongs asymptotically
to a $\chi^2$ distribution with two degrees of freedom.
We explain this
by the fact that once the cross section
$\sigma$ is set to zero the
likelihood function $L$ no longer depends on $m_W$ (all the $S_0$ and $S_m$ functions vanish)
and this is equivalent to fixing both the parameters of the fit at the same time.

In presence of the signal
the minimization procedure of $y$ described in the previous
section may be carried out semi--analytically using
Eq.(\ref{eq:chi2}).
One finds for $\sigma$
the estimator $\sigma_{est}$
given by:
\begin{eqnarray}\label{analitic}
  \sigma_{est}&=&\frac{\sum_{k}\frac{X_k\hat{S}_{m,k}(m_W^0)}{Var(X_k)}}{\sum_{k}
  \frac{\hat{S}_{m,k}(m_W^0)^2}
  {Var(X_{k})}}\\
  Var(\sigma_{est})&=&\left(\sum_{k}\frac{\hat{S}_{m,k}(m_W^0)^2}{Var(X_{k})}\right)
  ^{-1}.\label{analitic2}
\end{eqnarray}
\noindent
$\sigma_{\rm est}$ is a function of the data, such that
$<\sigma_{\rm est}>=\sigma$ (where brackets indicate mean value).
In Eqs.(\ref{analitic},\ref{analitic2}) $\hat{S}_{m,k}(m_W)\equiv
S_{m,k}(\sigma,m_W)/\sigma$ and $m_W^0$ is the WIMP mass that
maximizes the function $\delta(m_W)$,
which is given by:
\begin{equation}\label{eq:ratio}
  \delta(m_W)=
  \frac{\sum_{k}\frac{X_k\hat{S}_{m,k}(m_W)}{Var(X_{k})}}{\sqrt{\sum_{k}
  \frac{\hat{S}_{m,k}(m_W)^2}{Var(X_{k})}}}=\frac{\sigma_{est}}{\sqrt{Var(\sigma_{est})}}.
  \label{delta0}
\end{equation}
From equation (\ref{delta0})
$\delta\equiv\delta(m_W^0)$ may be interpreted, as expected,  as the
number of standard deviations of the signal, with:
\begin{eqnarray}\label{eq:vardelta}
  Var(\delta)&=& 1;\\
  <\delta>=<\frac{\sigma_{est}}{\sqrt{Var(\sigma_{est})}}>&
  =&\sqrt{<\delta^2>-1}\label{eq:meandelta}
\end{eqnarray}
\noindent so in presence of modulation
$\delta^2$ has the asymptotic distribution of a non central $\chi^2$ with one
degree of freedom and non centrality parameter given by
$(<\sigma_{\rm est}/\sqrt{Var(\sigma_{est})}>)^2=$$<\delta^2>-1$.

Using Eqs.(\ref{eq:chi2},\ref{eq:asimptotic}) we obtain:
\begin{equation}\label{expansion1}
  <\delta^2>=\frac{1}{2}\sum_{k}
  \frac{S_{m,k}(\sigma,m_W)^2\Delta
  E_k}{b_{k}+S_{0,k}}MT\alpha+2\label{eq:magic}.
\end{equation}
\noindent
where the same days of data taking have been assumed
for all the energy bins, and in the expression of $Var(X_k)$ the
following approximations have been made:
\begin{eqnarray}\label{eq:varx}
<\sum_i N_{ik} \cos^2\omega t_i>&\simeq& <N_{ik}>\sum_i\cos^2\omega t_i\\
<N_{ik}>&\simeq& W_k (b_k+S_0)
\end{eqnarray}
\noindent while we define the factor of merit
$\alpha\equiv \frac{2}{T}\sum_i \cos^2 \omega t_i$ ($\alpha$=1
in case of a full period of data taking)
and the terms depending on the $\beta_k$ have been neglected.

Equation (\ref{eq:magic}) allows to estimate the needed
exposure MT$\alpha$ in order to obtain a given value of $<\delta^2>$.
(It is worth noticing that the background $b_k$ indicates here the
amount of counts expected from radioactive contamination and noise, for
instance evaluated by making use of a Montecarlo simulation, and
not the average counts, that contain also the signal).
Whenever
the distribution of $\delta^2$ for a given WIMP model is
sufficiently far apart from that of the background, a given experimental
result may be discriminated to belong to one population or the
other. Once a
required $<\delta^2>$ is chosen, a sensitivity plot may be obtained by
showing the curves of constant MT$\alpha$ in the plane $m_W$--$\sigma$.

The statistical interpretation of the sensitivity plot is obtained
from the degree of overlapping between the distributions of $\delta^2$
in the two cases of absence and presence of modulation. The situation is
summarized in Table \ref{tab:cl}, where $<\delta^2>$ is tabulated as a function
of the fraction of experiments where absence of modulation can be
excluded (rows), and the required Confidence Level (columns).
For instance, if $<\delta^2>$=14.9,
there is a 90\% probability
to measure a value of $\delta^2$
higher than 7: this range would exclude the absence
of modulation at least at the 95º\% C.L. In the same way
a less demanding
value of $<\delta^2>$=5.6 would give a 50\% probability to see an
effect at the 90\% C.L. or more. In the following section, we
will adopt these two representative values in order to discuss the
prospects of modulation searches.
This procedure allows to
see in a compact way the masses and exposures needed to explore each
different region of the WIMP $m_W$--$\sigma$ parameter space.

The expression of $<\delta^2>$ in Eq.(\ref{eq:magic}) deserves a short
comment. Some attention has been devoted in discussions on the perspectives
of modulation searches\cite{hasenb,ramachers}
to the fact that since the $S_{m,k}$ parameters can vanish or be negative,
cancellations may arise in the calculation of the
signal--to--noise ratio, so that an optimal choice should be found
for the energy threshold or for the amplitude of the energy interval where
the signal is integrated.
According to
Eq.(\ref{eq:magic}), 
in order to exploit all the information available, the signal--to--noise ratio should be
calculated over the whole energy spectrum, using the smallest bin width
allowed by statistics and
resolution, and then combined quadratically, i.e.:
\begin{equation}\label{inverses}
     ({\rm signal/noise})^2_{{\rm combined}}=\sum_{i} ({\rm signal/noise})^2_i
\end{equation}
\noindent where $i$ indicates the energy bin. Of course
in this procedure it is crucial that the count
rates of different bins are not correlated.

\section{Sensitivity plots and quantitative discussion}

In this section we give a quantitative discussion of the sensitivity plots
by considering
different target materials: $Ge$, $TeO_2$, $NaI$ with quenching factors,
experimental thresholds and
resolutions summarized in Table \ref{tab:parameters} (they have
been chosen in order to be indicative of running or planned
experiments\cite{dama,heid99,igex,cdms,cuore}).
For simplicity, in order to calculate the
sensitivity plots, a flat background will be always assumed and in the
following the index $k$ of $b_k$ will be dropped.

As far as the astrophysical parameters entering in the calculation are concerned,
we assume as usual that the WIMP velocities in the halo follow a Maxwellian
distribution. The WIMP r.m.s. velocity $v_{\rm rms}$ is
related to the measured rotational velocity of
the Local System at the Earth's position $v_{\rm loc}$ by the relation
$v_{\rm rms}^2=\frac{2}{3} \; v_{\rm loc}^2$, while
the Sun's velocity $v_{\rm sun}$ in the galactic rest frame
is given by $v_{\rm sun}\simeq (v_{\rm loc}+12)$ km
sec$^{-1}$ (here the motion of the solar system with respect to the Local System is taken
into account) and the Earth velocity is:
\begin{equation}\label{eq:earthvel}
  v_{\rm earth}=v_{\rm sun}+v_{\rm orb}
  \sin\delta\cos\left[\omega(t-t_0)\right]
\end{equation}
\noindent where $v_{orb}\simeq$ 30 km sec$^{-1}$ and
$\sin\delta\simeq$ 0.51\cite{frees} ($\delta$ is the angle between the Ecliptic
and the Galactic plane).

As it has been pointed out in the literature\cite{velocities} the present uncertainty
on $v_{\rm loc}$ can affect the final result of WIMP direct detection
calculations in a significant way.
In the following we will vary $v_{\rm loc}$ in its
physical range, $v_{\rm loc}=(220 \pm 50)$ km sec$^{-1}$.
Finally, we adopt an escape velocity
$v_{\rm esc}$=650 km/sec and take for the local halo mass density the value
$\rho$=0.3 GeV/cm$^3$.

In order to compare the results for different target materials in the same
plots, we show our results in the plane $m_W-\sigma^{(n)}$, where $\sigma^{(n)}$
is the WIMP cross section rescaled to the nucleon by adopting a
scalar--type interaction (such as the one that would be dominant
in the case of a neutralino). In the case of a monoatomic target this implies the
transformation:
\begin{equation}\label{cross}
  \sigma=\sigma^{(n)} A^2 \frac{\mu^2_{W,N}}{\mu^2_{W,n}}
\end{equation}
\noindent where $A$ is the target atomic number, $\mu_{W,N}$ is the
WIMP--nucleus reduced mass and $\mu_{W,n}$ the
WIMP--nucleon reduced mass. Generalization to a multi--target
species is straightforward.

In Figures 1--5
we give an estimate of the minimal
exposures needed to explore the WIMP parameter space by calculating the sensitivity plots
for $<\delta^2>$=5.6 and $v_{\rm loc}$=220 km sec$^{-1}$.
In all the figures the curves are obtained for
values of MT$\alpha$ ranging from 10 kg$\cdot$ year to 100 kg$\cdot$ year
in steps of 10 (from top to bottom). The closed contour and the cross indicate
respectively
the 2$\sigma$ C.L. region singled out by the DAMA modulation search experiment and the minimum
of the likelihood function found by the same authors\cite{dama}.
Note that by Eq.(\ref{eq:magic}) each sensitivity plot corresponds to a fixed value of the
quantity $(<\delta^2>-2)/{\rm MT}\alpha$, so the curves may be easily
rescaled to different values of $<\delta^2>$ by a suitable change in
the corresponding exposures.
As expected, the
sensitivity is roughly proportional to $({\rm MT})^{-1/2}$.

Figures 1--2
refer to Ge detectors with background b=0.01 cpd/kg/keV and
energy thresholds E$_{th}$=2
and 12 keV respectively (thresholds already obtained in the
COSME\cite{cosme} and Heidelberg--Moscow\cite{heid99} experiments,
respectively, and a background not far from that obtained by
Heidelberg--Moscow and CDMS\cite{cdms}).
In Figure 3
the intermediate situation of a germanium
detector with b=0.1 cpd/kg/keV and E$_{th}$=4 keV is given
(except from the assumption of flat background, the case depicted
in Fig. 3 corresponds to the present performances of the IGEX
experiment\cite{igex}).
The example of a TeO$_2$ detector is shown in Figure 4
(with b=0.01 cpd/kg/keV and
E$_{th}$=5 keV, which are the foreseen performances of
CUORICINO\cite{cuore}) while an NaI detector with
b=0.1 cpd/kg/keV and E$_{th}$=2 keV
(practically the DAMA\cite{dama} values)
is depicted in Figure 5.
We recall once more that the values of $b$ quoted for the
aforementioned experiments are the levels of counts only due to the background,
i.e. they do not include the possible signal.

In order to be sensitive to a possible halo WIMP,
sensitivity plots need to lie below the current most stringent
upper limits to the WIMP--nucleon scalar cross section
$\sigma^{(n)}$. Figures 6--9
show the corresponding required minimal exposures
as a function of the WIMP mass $m_W$ for the cases of
Ge, TeO$_2$ and NaI and calculated for the exclusion plot of
Ref.\cite{damlim}, obtained with data statistically discriminated by pulse shape analysis.
In each plot, dotted, solid and dashed curves refer to $v_{\rm loc}$= 170, 220 and
270 km sec$^{-1}$ respectively, and in each case
the different examples of b=0,0.01 and 0.1 cpd/kg/keV are given from bottom to
top.

The effect of the uncertainty on $v_{\rm loc}$ is to displace horizontally the
sensitivity plots to higher WIMP masses for
lower values of $v_{\rm loc}$. This implies that for a
given value of MT the upper limit on the explorable WIMP mass can
significantly rise if $v_{\rm loc}$ is taken in its lowest range.
On the other hand, the curves of Figs.6--9
do not depend on $\rho_{\rm loc}$, and
are almost insensitive to $v_{\rm
esc}$, unless for very low WIMP masses.

The results of the present discussion are summarized
in tables \ref{tab:summary220}--\ref{tab:summary270}.
In order to show the dependence of theoretical expectations on $v_{\rm
loc}$, in the three tables the values $v_{\rm loc}$=220, 170 and 270 km
sec$^{-1}$ are used respectively.
In the tables we add examples with other values of the energy
thresholds and the background b (the complete set is given in table
\ref{tab:parameters}).

In the second column of tables \ref{tab:summary220}-\ref{tab:summary270}
the minimal MT$\alpha$
values that are necessary to explore the regions of the $m_W$--$\sigma^{(n)}$ plane
below the
exclusion plot of Ref.\cite{damlim} are given for  $<\delta^2>=5.6$, followed, if necessary,
by a corresponding interval
for the WIMP mass. If no interval is shown, the given exposure allows to reach
the exclusion plot over the whole considered range 10 GeV$<m_W<$1000
GeV.

In the last two columns of tables
\ref{tab:summary220}-\ref{tab:summary270}
we summarize the prospects of exploration of the DAMA
region. The values of MT$\alpha$ given in the third column
correspond to the lowest values that give a $<\delta^2>$=5.6 sensitivity plot encompassing
all the 2$\sigma$ contour. Finally, in the last column the same values
are given in the case of $<\delta^2>$=15.

\section{Conclusions}

In the present paper the prospects for direct searches of a WIMP modulation effect
are discussed in the case of different target materials and
as a function of the detector mass M and the time of exposure T.

Given its purely statistical nature, such a discussion
has not taken into account the problems related to the systematics of a real
experiment, exacerbated by the challenge of detecting a small
signal depending on time. Consequently, although the
exposure MT planned or reachable in a projected experiment
can be an indication of its future chances,
the real prospects of a given modulation search rely in
no lesser extent in its realistic possibilities of
keeping background stability under control for very long periods of time.
For instance,
in the examples previously discussed the signal typically amounts to a fraction
between 1\%
and 5\% of the average count rates, concentrated in the low--energy range of the spectrum.
This implies that the corresponding WIMP
models would be explorable only
with a control of the stability of the background
substantially below that range.

We also remark that in our calculation a given sensitivity is
associated to a value of the product MT$\alpha$,
irrespective to the actual number of the accumulated
periods. This is due to the fact that
the signal/noise ratio discussed in the text scales as $1/\sqrt{N}$
with $N$ the total amount of counts in case of Poissonian fluctuations
(the contribution of data collected in the days near the maximum or the minimum of the signal
is enhanced by the factor of merit $\alpha$).
However in a real experiment the collection of several periods
would be preferred in order to have a
higher control of systematics effects.

An important feature of all the plots is that the
sensitivity to modulation is generally a decreasing function of
the WIMP mass, the highest sensitivities corresponding roughly
to the interval $10 \;{\rm GeV}\lsim m_W \lsim 130\;{\rm GeV}$.
However, the actual upper limit of the region of WIMP masses within the
reach of a given experiment depends in a crucial way from the
choice of the astrophysical parameters, something already
discussed in the literature \cite{velocities}.

The optimal
energy intervals where the signal should be looked for can be found for instance by requiring that
the sensitivity plot does not change more than 5\% over the whole
range of $M_W$ when more energy bins are added. With such proviso we
find
$2 \div 10$ keV$\lsim E\lsim 35$ keV for Ge,
$2 \div 5$ keV$\lsim E\lsim 30$ keV for TeO$_2$ and
$2$ keV$\lsim E\lsim 8$ keV for NaI, the
lower bounds depending on the WIMP mass.

The region singled out by the DAMA experiment is within the reach of many realistic set-ups.
This can be quantified by looking at the Figs.1--5 (for $v_{\rm loc}$=220 km sec$^{-1}$)
and is shown in tables \ref{tab:summary220}--\ref{tab:summary270} (also for $v_{\rm
loc}$=170 km sec$^{-1}$ and 270 km sec$^{-1}$).
For instance, for $v_{\rm loc}$=220 km sec$^{-1}$, the typical needed
exposure
for a germanium detector turns up to be MT$\alpha\simeq$ 50
kg$\cdot$year in the two realistic examples of $E_{\rm th}$=2--4 keV, b=0.1
cpd/kg/keV and $E_{\rm th}$=12 keV, b=0.01 cpd/kg/keV, and
lowers to about MT$\alpha\simeq$ 15 kg$\cdot$year in the more optimistic case of
$E_{\rm th}$=2-4 keV and b=0.01 cpd/kg/keV.

However, as can be seen from Tables \ref{tab:summary170} and \ref{tab:summary270},
for all the considered nuclear targets
these estimations depend in a critical way
on the value of $v_{\rm loc}$,
detection chances being systematically better for lower values.

As far as TeO$_2$ is concerned, the achievable sensitivities seem
promising over the whole considered WIMP mass interval 10 GeV$\lsim m_W\lsim$ 1000 GeV,
and an exposure as low as $\simeq$ 25 kg$\cdot$year could be sufficient to start
a modulation search exploration if b=0.01 and E$_{th}$=2 keV.
However, when comparing these results with other kinds of detector,
it is worth noticing that these sensitivities are very
dependent on the background and threshold, and
our assumptions are to be considered as future goals
since high mass TeO$_2$ bolometers are still at the R\&D stage of
development.

Turning to NaI detectors, they are the first that have been used to
search for modulation\cite{modza,dama}, essentially due to the
possibility to reach high masses. Their prospects for detection are
a sensitive function of the obtainable threshold, since the most
part of the signal is contained in the first energy bins, $E\lsim
5$ keV, because of the low quenching factor. As can be seen from
Fig.9, their sensitivity is higher for relatively light WIMP's,
and
the minimal required exposures turn out to be very steep functions
of the WIMP mass, the heavier explorable values of $m_W$ depending
critically on $v_{\rm loc}$ and reaching up to $m_W\simeq$ 130 GeV.

The experimental region singled out by Ref.\cite{dama} (30 GeV$\lsim m_W \lsim$ 130 GeV
for $v_{\rm loc}$=220 km sec$^{-1}$) extends well beyond the upper limit for the detectable
WIMP masses implied by the curves of Fig.9 ($m_W\lsim 70$ GeV for the same value of
$v_{\rm loc}$). An explanation of this fact is that the experimental region encompasses
configurations with much lower probabilities (corresponding to a 2--$\sigma$ Confidence Level)
than the ones required in the sensitivity plots.
For $m_W$ and $\sigma^{(n)}$
corresponding to the minimum of Ref.\cite{dama} and assuming MT$\alpha$=50 kg year,
$b$=0.5 cpd/kg/keV and $E_{\rm th}=2$ keV, we find $<\delta^2>\simeq$
7.2. The experimental result published
by Ref.\cite{dama} is $\delta^2=8.23$.

The overall picture arising from the examination
of Figs. 1--5 and the results tabulated in Tables \ref{tab:summary220}--\ref{tab:summary270}
turns out to be dependent, as already mentioned, on the choice of the astrophysical parameter
$v_{\rm loc}$. However a general trend can be drawn, in which
prospects of modulation
searches seem promising provided that the WIMP signal
is not far below present sensitivities.
The lowest values of
explorable $\sigma^{(n)}$ fall in most cases in the typical range of
few $\times 10^{-10}$ nbarn.  This value can lower up to one order of magnitude
in some extreme cases (for instance, for MT$\alpha$=1000
kg$\cdot$year, a low threshold of a few keV and setting b=0).

The impact that future improvements on
exclusion plots would have on the prospects of modulation searches can be estimated in the
following way. In Eq.(\ref{eq:magic}), at fixed $<\delta^2>$, when in the denominators
the inequality $b_k>S_{0,k}$ holds, MT$\alpha$ is (approximately) proportional to $\sigma^2$.
On the other hand, when $b_k<S_{0,k}$, MT$\alpha$ is proportional to $\sigma$.
This implies that an improvement of one order of magnitude of the exclusion plot
would rise the curves of Figs. 6--9
between one and two orders of magnitude,
depending on the background achieved by the modulation search.

As a last remark, we make a few comments
about the case of a null result, when
modulation can be used for background subtraction, improving so the exclusion plot one
would obtain from time--integrated counts.
Numerical results show that the effectiveness of this technique
improves the exclusion plot only in case of relatively high count-rate
levels. This is confirmed by numerical inspection of
Eq.(\ref{eq:magic}):
in order for a modulation search experiment with background
$b_1$ to extract a better exclusion plot than another experiment
with background $b_2$, the following approximate
condition must hold
\begin{equation}\label{eq:limsup}
  \frac{\sqrt{b_1}}{b_2}\lsim \sqrt{\frac{MT/{(\rm kg \;year)}}{2(\delta^2_{\rm
  lim}-2)}} F(m_W)\;\; ; \;\;F(m_W)\equiv \frac{\sqrt{\sum_k S_{\rm m,k}^2\Delta
  E_k}}{S_{0,threshold}},
\end{equation}
\noindent where $b_1$ and $b_2$ are in cpd/kg/keV, $\delta^2_{\rm lim}$
is the upper limit to $<\delta^2>$ provided by the null modulation
search, the $S_{\rm m,k}$'s refer to experiment 1,
and $S_{\rm 0,threshold}$ refers to experiment 2, assuming
that its upper limit is driven by its background at threshold.
Typically, at the 90\% C.L., one
can assume $\delta_{\rm lim}\simeq 10$, while a numerical check in
experimental situations analogous to the ones analyzed in the previous sections gives
$F(m_W)\lsim 0.1$ and usually much lower.
In this case Eq.(\ref{eq:limsup}) implies
$b_1\lsim b_2^2\times 0.1\times$ MT,
that, for instance, for $b_1=b_2\equiv b$ and MT=100 kg year
implies that a modulation analysis can improve the exclusion plot only if
$b\gsim 0.1$ cpd/kg/keV.

\vspace{1cm}
\section { Acknowledgements}
This search has been partially supported by the Spanish Agency of
Science and Technology (CICYT) under the grant AEN99--1033 and the
European Commission (DGXII) under contract ERB-FMRX-CT-98-0167.
One of us (S.S.) acknowledges the partial
support of a fellowship of the INFN (Italy).


\begin{table}
\caption{Parameter $<\delta^2>$ as a function of the required Confidence Level
(columns) and the fraction of successful experiments(rows).\label{tab:cl}}
\begin{center}
\begin{tabular}{|c|c|c|c|c|c|} \hline
 & 90\% & 95\% &  99\% &  99.5\% & 99.9\%  \\ \hline
 50\%      & 5.6  & 7.0 & 10.2  & 11.6 & 14.8 \\ \hline
 60\%      & 6.8  & 8.3 & 11.8  & 13.3 & 16.8 \\ \hline
 70\%      & 8.1  & 9.9 & 13.7  & 15.3 & 19.0 \\ \hline
 80\%      & 9.9  & 11.8 & 16.0 & 17.8 & 21.8 \\ \hline
 90\%      & 12.8  & 14.9 & 19.6 & 21.6 & 26.0 \\ \hline
 95\%      & 15.4  & 17.8 & 22.9 & 25.0 & 29.8 \\ \hline
 99\%      & 21.0  & 23.8 & 29.7 & 32.2 & 37.5 \\ \hline
\end{tabular}
\end{center}
\end{table}

\begin{table}
\caption{Summary of the experimental parameters
assumed in WIMP modulation searches for the different target
materials discussed in the text. They are inspired to the running or
projected experiments whose reference is shown in the last
column.
\label{tab:parameters}}
\begin{center}
\begin{tabular}{|c|c|c|c|c|c|} \hline
& & $E_{th}$ (keV) & FWHM(keV)& Quenching & Ref.\\ \hline
& Ge     & 2;4;12  & 1 & 0.25 &\cite{heid99,igex,cdms} \\ \hline
& NaI     & 2 & 2 & 0.09 &\cite{dama}   \\ \hline
& TeO$_2$  & 2;5& 2 & 0.93 & \cite{cuore} \\ \hline
\end{tabular}
\end{center}
\end{table}

\begin{table}
\caption{ Summary of minimal exposures, all in kg $\cdot$ year,
for $v_{\rm loc}=220$ km sec$^{-1}$. E$_{th}$ indicates the energy
thresholds expressed in keV, b the background (assumed constant in energy) in
cpd/kg/keV. Exposures are estimated for the WIMP mass range in
parenthesis if shown, otherwise over the range 10$\lsim
m_W\lsim$1000.\label{tab:summary220}}
\begin{center}
\begin{tabular}{cccc} \hline
                     & Exploration of not       & DAMA region:      & DAMA region:     \\
                     & excluded regions ($<\delta^2>=5.6)$ & $<\delta^2>=5.6$  & $<\delta^2>=15$  \\ \hline
Ge                   &                          &                   &                  \\
E$_{th}$=2,  b=0.01  &      35                  &    15             &      55          \\
                     &20 (m$_W\lsim$110)        &                   &                  \\
E$_{th}$=2, b=0.1    &        80                &    50             &     175          \\
                     &55 (m$_W\lsim$110)        &                   &                 \\
E$_{th}$=12, b=0.01  &25 (40$\lsim m_W\lsim$110)&     50           &     190          \\
E$_{th}$=12, b=0.1   &100 (45$\lsim m_W\lsim$110)&   330            &    1190          \\
E$_{th}$=4,  b=0.1   &           155            &     50            &     190          \\
                     &     65 (m$_W\lsim$110)   &                   &                  \\ \hline
TeO$_2$              &                          &                   &                  \\
E$_{th}$=2,  b=0.01  &        20                &     25            &      90          \\
E$_{th}$=2, b=0.1    &        40                &     55            &      210          \\
E$_{th}$=5, b=0.01   &        40                &     40            &     150          \\
E$_{th}$=5, b=0.1    &        85                &     120           &     435          \\ \hline
NaI                  &                          &                   &                  \\
E$_{th}$=2,  b=0.1   &        50(m$_W\lsim$70)   &     180          &      660         \\
\end{tabular}
\end{center}
\end{table}

\begin{table}
\caption{The same as in Table \protect\ref{tab:summary220}
for $v_{\rm loc}$=170 km sec$^{-1}$.
\label{tab:summary170}}
\begin{center}
\begin{tabular}{cccc} \hline
                     & Exploration of not       & DAMA region:      & DAMA region:     \\
                     & excluded regions ($<\delta^2>=5.6)$ & $<\delta^2>=5.6$  & $<\delta^2>=15$  \\ \hline
Ge                   &                          &                  &                  \\
E$_{th}$=2,  b=0.01  &      20                  &    8             &      30          \\
                     &6 (m$_W\lsim$110)         &                  &                  \\
E$_{th}$=2, b=0.1    &        50                &    25            &     90           \\
                     &20 (m$_W\lsim$110)        &                  &                  \\
E$_{th}$=12, b=0.01  &30 (m$_W\gsim$45)         &     25           &     95           \\
E$_{th}$=12, b=0.1   &135 (m$_W\gsim$35)        &   160            &    570           \\
E$_{th}$=4,  b=0.1   &           90             &     25           &     90           \\
                     &20 (m$_W\lsim$110)        &                  &                  \\ \hline
TeO$_2$              &                          &                  &                  \\
E$_{th}$=2,  b=0.01  &        15                &     15           &      50          \\
E$_{th}$=2, b=0.1    &        25                &     30           &      110         \\
E$_{th}$=5, b=0.01   &        25                &     20           &     80           \\
E$_{th}$=5, b=0.1    &        50                &     65           &     230          \\ \hline
NaI                  &                          &                  &                  \\
E$_{th}$=2,  b=0.1   &        50(m$_W\lsim$125) &     100          &      355         \\
\end{tabular}
\end{center}
\end{table}

\begin{table}
\caption{
The same as in Table \protect\ref{tab:summary220}
for $v_{\rm loc}$=270 km sec$^{-1}$.\label{tab:summary270}}
\begin{center}
\begin{tabular}{cccc} \hline
                     & Exploration of not       & DAMA region:      & DAMA region:     \\
                     & excluded regions ($<\delta^2>=5.6)$ & $<\delta^2>=5.6$  & $<\delta^2>=15$  \\ \hline
Ge                   &                           &                  &                  \\
E$_{th}$=2,  b=0.01  &      50                   &    25            &      90          \\
                     &40 ($m_W\lsim$110)         &                  &                  \\
E$_{th}$=2, b=0.1    &        105                &    80            &     285          \\
E$_{th}$=12, b=0.01  &70 (25$\lsim m_W\lsim$110) &    70            &     260          \\
E$_{th}$=12, b=0.1   &300 (25$\lsim m_W\lsim$110)&   480            &    1735          \\
E$_{th}$=4,  b=0.1   &           210             &    85            &     310          \\
                     &180 ($m_W\lsim$110)        &                  &                  \\ \hline
TeO$_2$              &                           &                  &                  \\
E$_{th}$=2,  b=0.01  &        30                 &    40            &     140          \\
E$_{th}$=2, b=0.1    &        60                 &    90            &     330          \\
E$_{th}$=5, b=0.01   &        60                 &    65            &     230          \\
E$_{th}$=5, b=0.1    &       125                 &   190            &     680          \\ \hline
NaI                  &                           &                  &                  \\
E$_{th}$=2,  b=0.1   & 180(m$_W\lsim$60, m$_W\gsim$200)&     305    &    1100          \\
\end{tabular}
\end{center}
\end{table}

\newpage

\begin{figure}[p]
\vspace{4.5 cm}
\begin{center}
\mbox{\psfig{figure=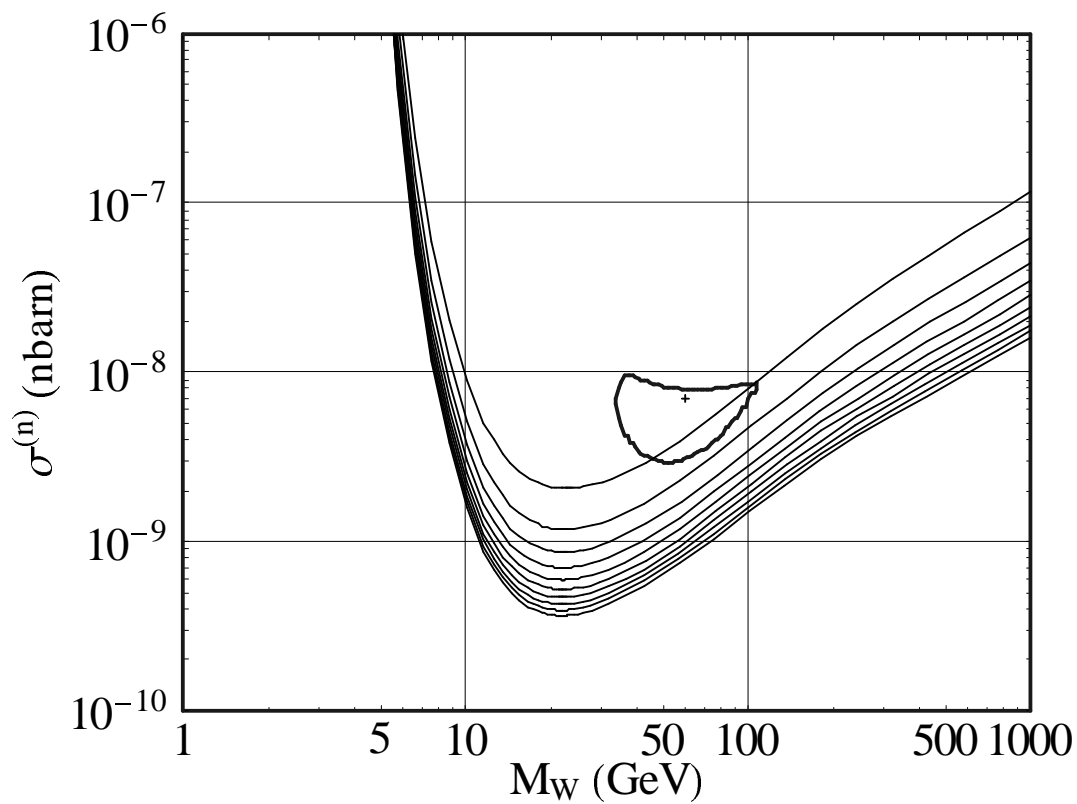,height=110mm}}
\caption{
Sensitivity plots in the $\sigma^{(n)}$--$m_W$ plane for
a germanium detector with threshold energy E$_{th}$=2 keV, flat
background b=0.01 cpd/kg/keV and calculated for
$<\delta^2>$=5.6. The set of curves correspond to different values
of the exposure, MT$\alpha$=10 to 100 kg$\cdot$year in steps of
10 from top to bottom. The closed contour represents the 2$\sigma$
C.L. region singled out by the modulation analysis performed by the
DAMA experiment\protect\cite{dama} and the cross indicates the minimum of the likelihood
found by the same authors.
\label{fig:ge_eth2_b001_mtg}}
\end{center}
\end{figure}

\newpage

\vspace*{5 cm}
\begin{figure}[p]
\begin{center}
\mbox{\psfig{figure=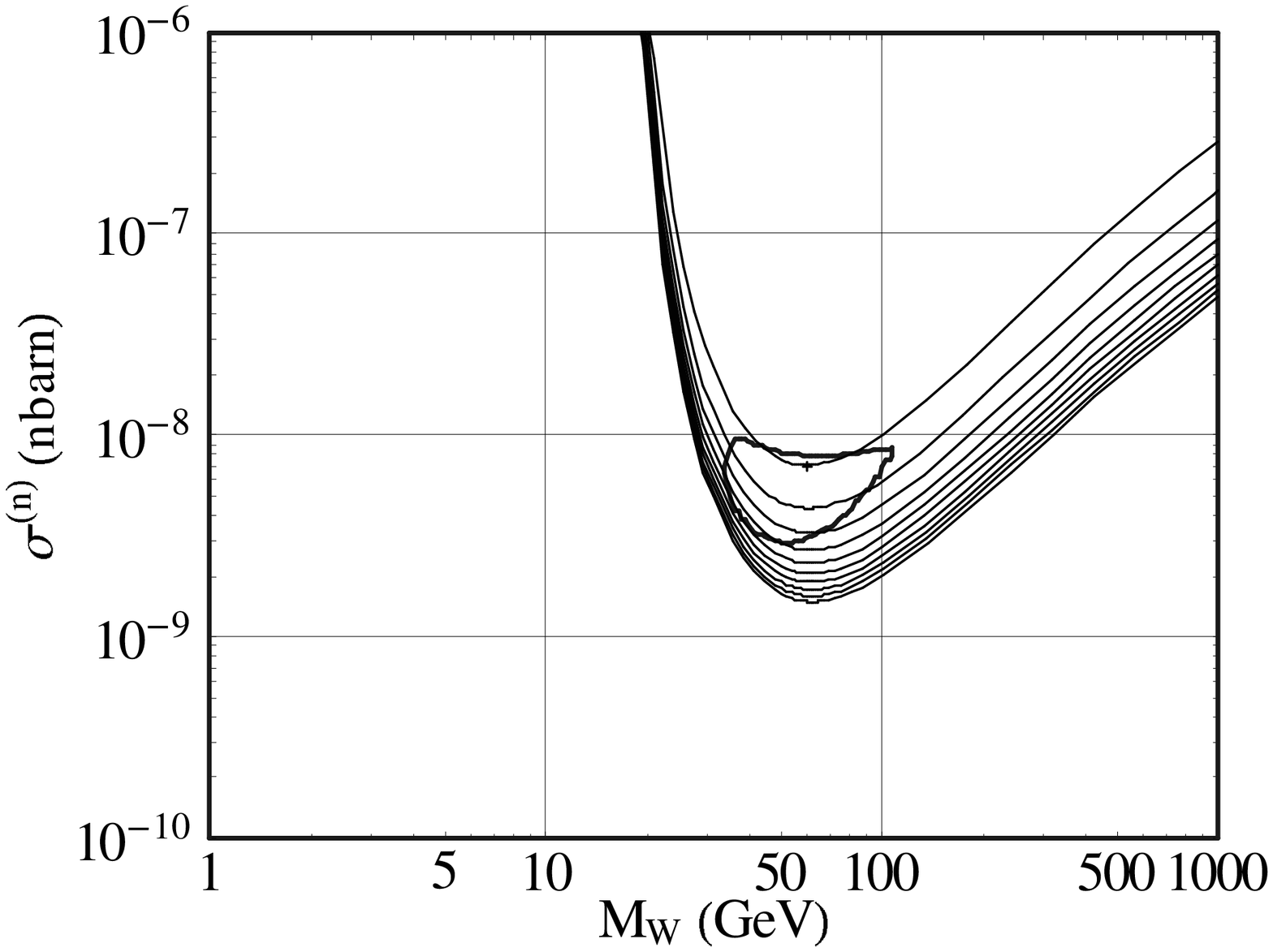,height=110mm}}
\caption{The same as in Figure 1 for
a germanium detector with threshold energy E$_{th}$=12 keV.
\label{fig:ge_eth12_b001_mtg}}
\end{center}
\end{figure}

\newpage

\vspace*{5 cm}

\begin{figure}[p]
\begin{center}
\mbox{\psfig{figure=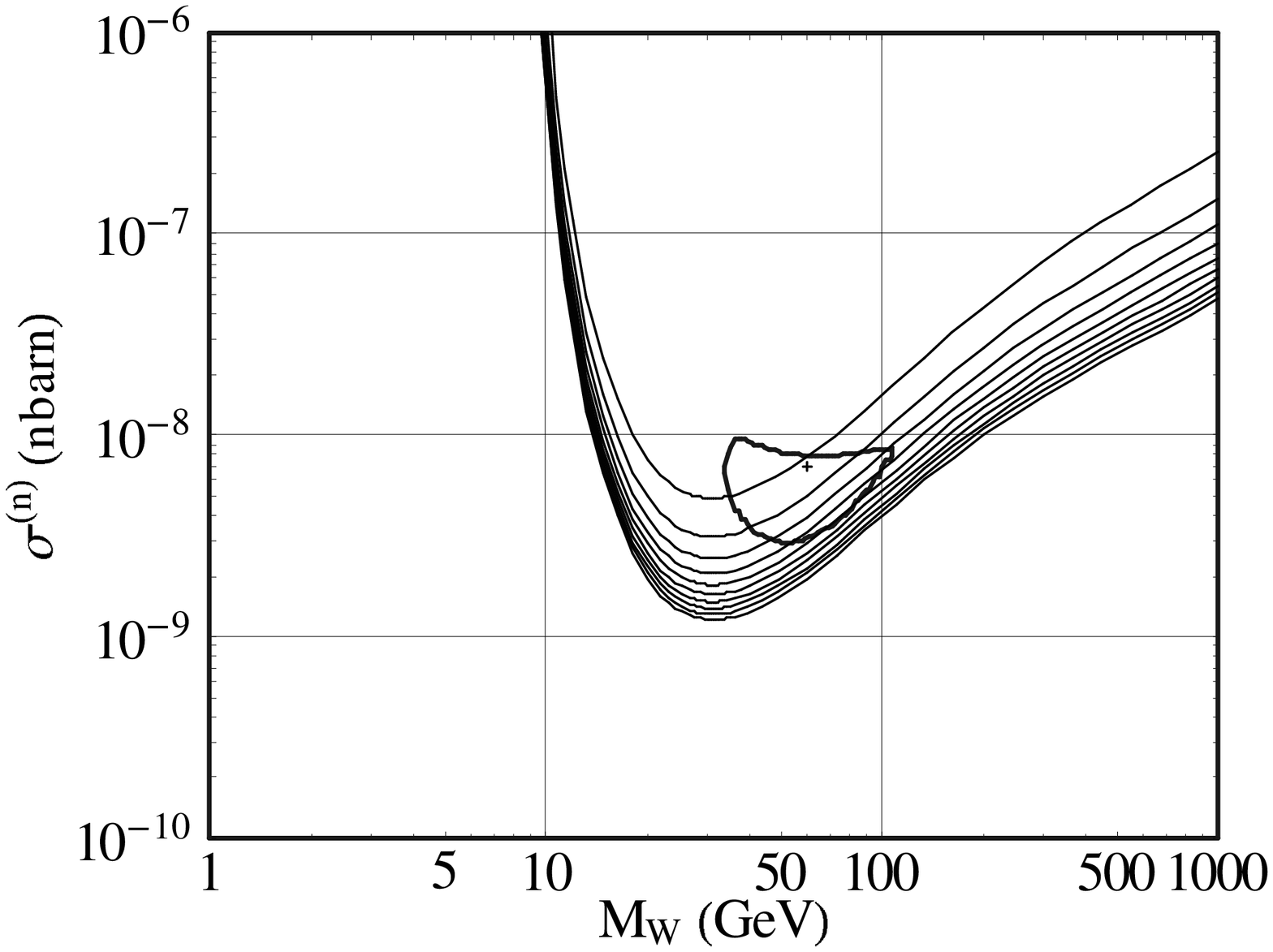,height=110mm}}
\caption{The same as in Figure 1 for
a germanium detector with threshold energy E$_{th}$=4 keV and flat
background b=0.1 cpd/kg/keV.
\label{fig:ge_eth4_b01_mtg}}
\end{center}
\end{figure}

\newpage

\vspace*{5 cm}

\begin{figure}[p]
\begin{center}
\mbox{\psfig{figure=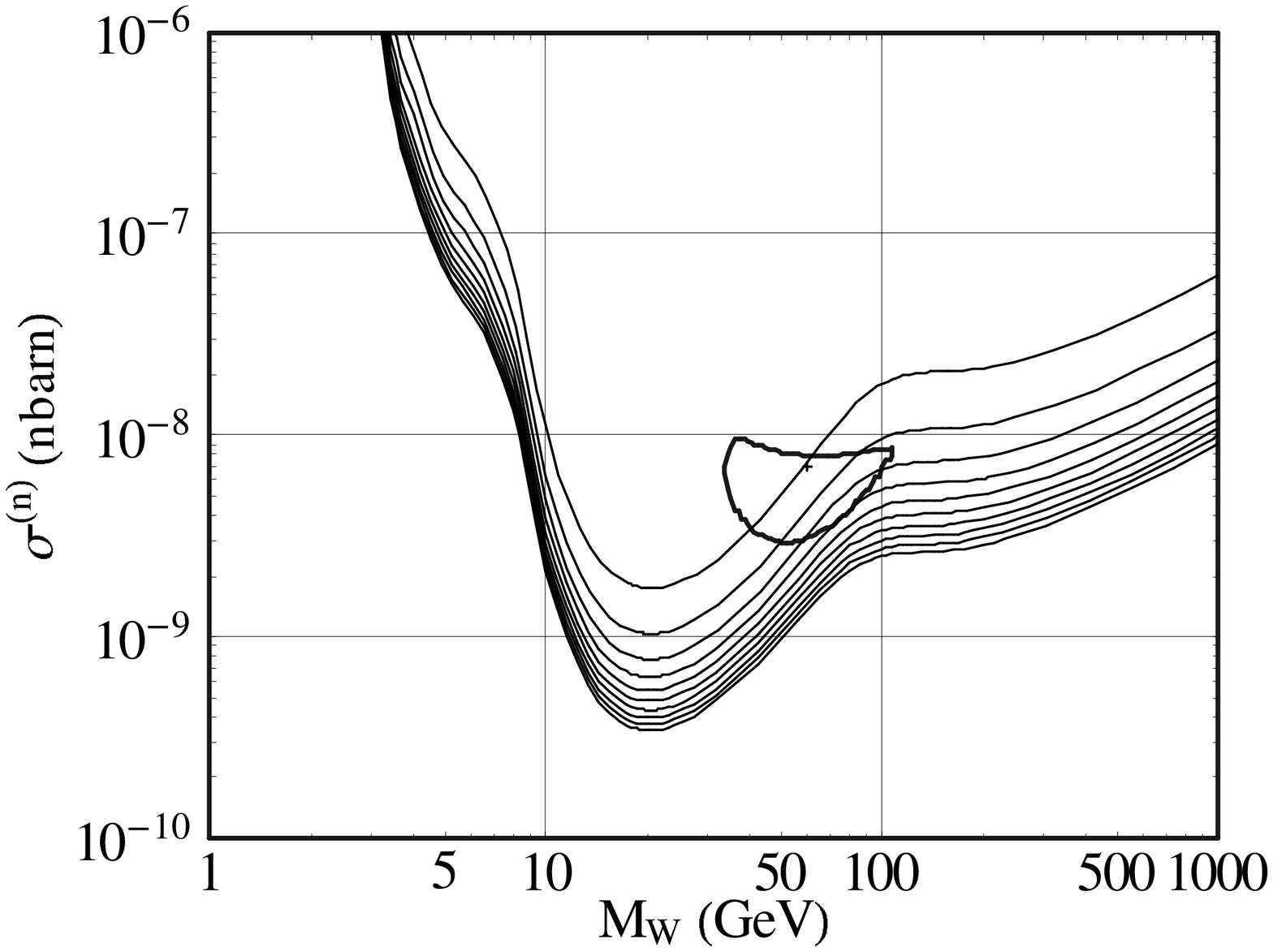,height=110mm}}
\caption{The same as in Figure 1 for
a TeO$_2$ detector with threshold energy E$_{th}$=5 keV and flat
background b=0.01 cpd/kg/keV.
\label{fig:teo2_eth5_b001_mtg}}
\end{center}
\end{figure}

\newpage

\vspace*{5 cm}

\begin{figure}[p]
\begin{center}
\mbox{\psfig{figure=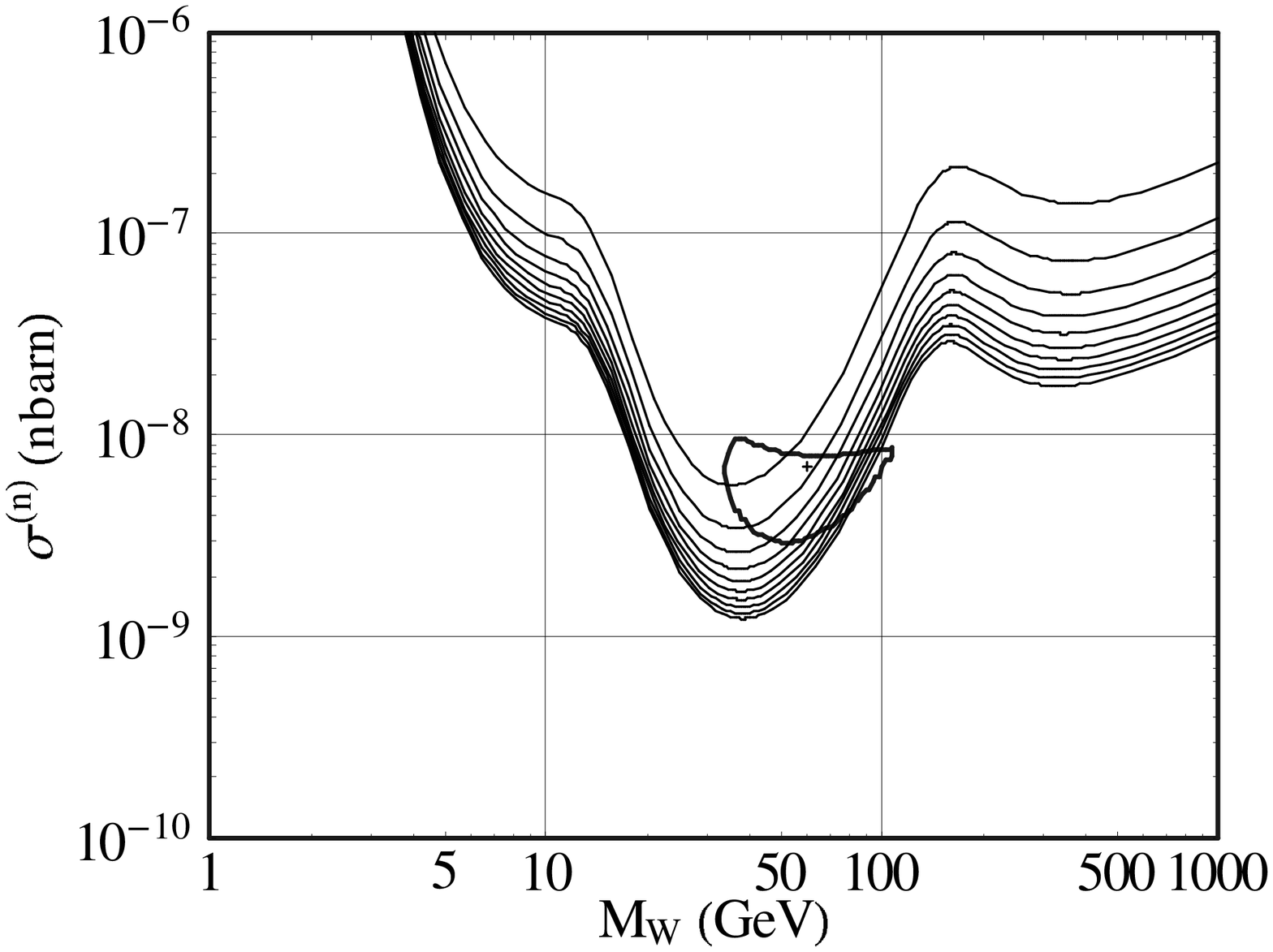,height=110mm}}
\caption{The same as in Figure 1 for
an NaI detector with threshold energy E$_{th}$=2 keV and flat
background b=0.1 cpd/kg/keV.
\label{fig:nai_eth2_b01_mtg}}
\end{center}
\end{figure}

\newpage

\vspace*{5 cm}

\begin{figure}[p]
\begin{center}
\mbox{\psfig{figure=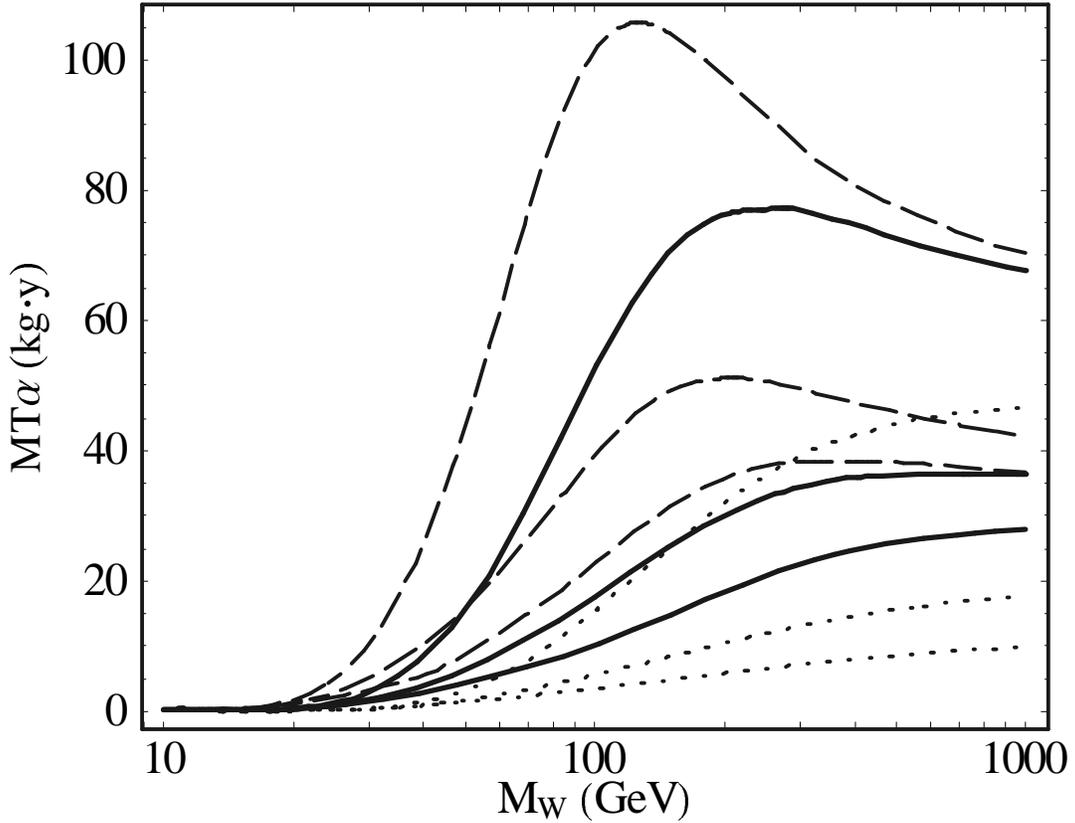,height=110mm}}
\caption{Minimal exposure MT$\alpha$ as a function of the WIMP mass
$m_W$ required for the $<\delta^2>$=5.6 sensitivity plot to reach the exclusion plot of
Ref.\protect\cite{dama} to the
WIMP--nucleon scalar cross section $\sigma^{(n)}$, and calculated
for a germanium
detector with threshold energy E$_{th}$=2 keV.
Dotted, solid and dashed curves refer to $v_{\rm loc}$= 170, 220 and
270 km sec$^{-1}$ respectively. For each value of $v_{\rm loc}$
the curves calculated for a flat background b=0,0.01 and 0.1 cpd/kg/keV
are given from bottom to top.
\label{fig:ge_eth2_mtmin}}
\end{center}
\end{figure}

\newpage

\vspace*{5 cm}

\begin{figure}[p]
\begin{center}
\mbox{\psfig{figure=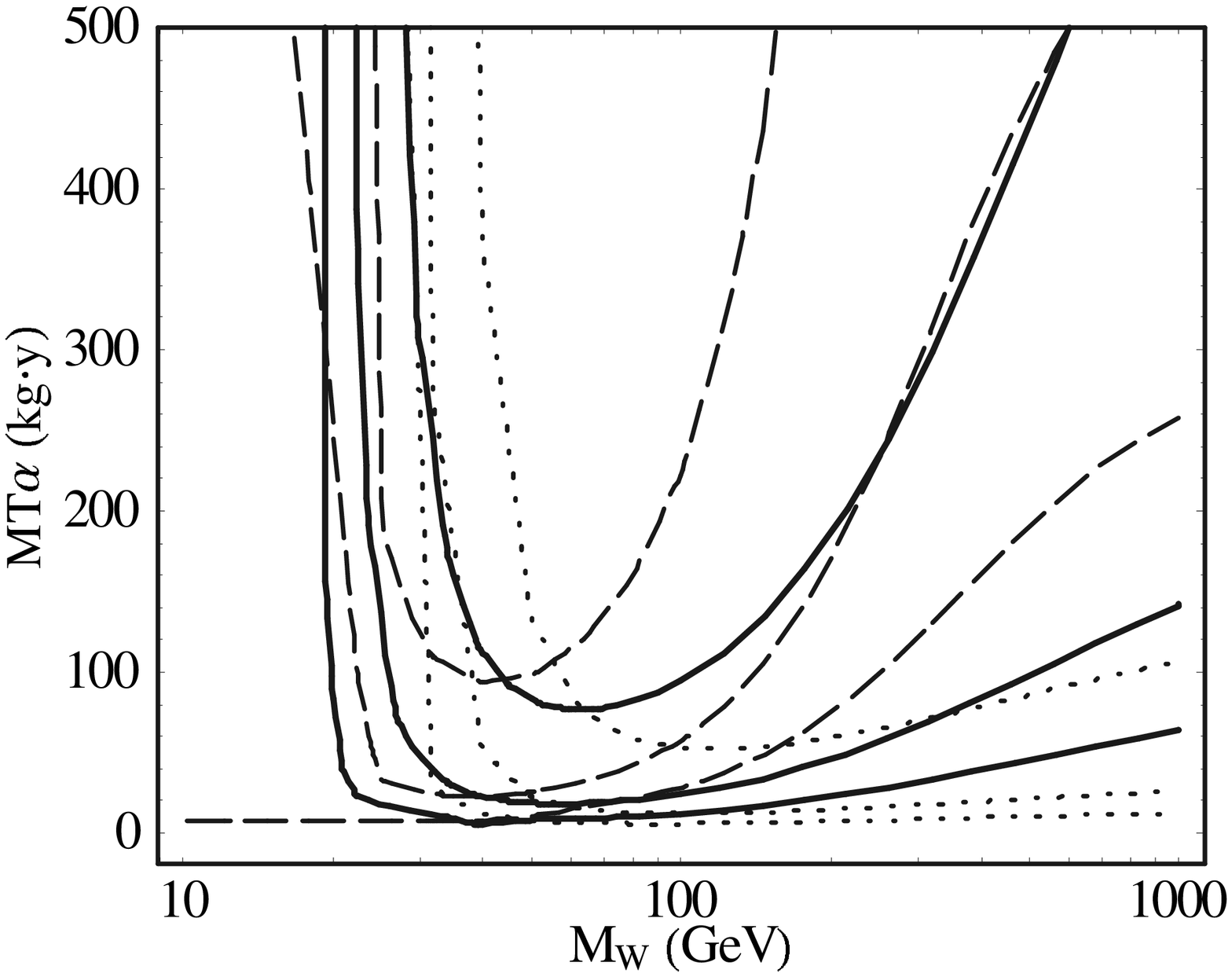,height=110mm}}
\caption{The same as in Figure 6
for a germanium
detector with threshold energy E$_{th}$=12 keV.
\label{fig:ge_eth12_mtmin}}
\end{center}
\end{figure}

\newpage

\vspace*{5 cm}

\begin{figure}[p]
\begin{center}
\mbox{\psfig{figure=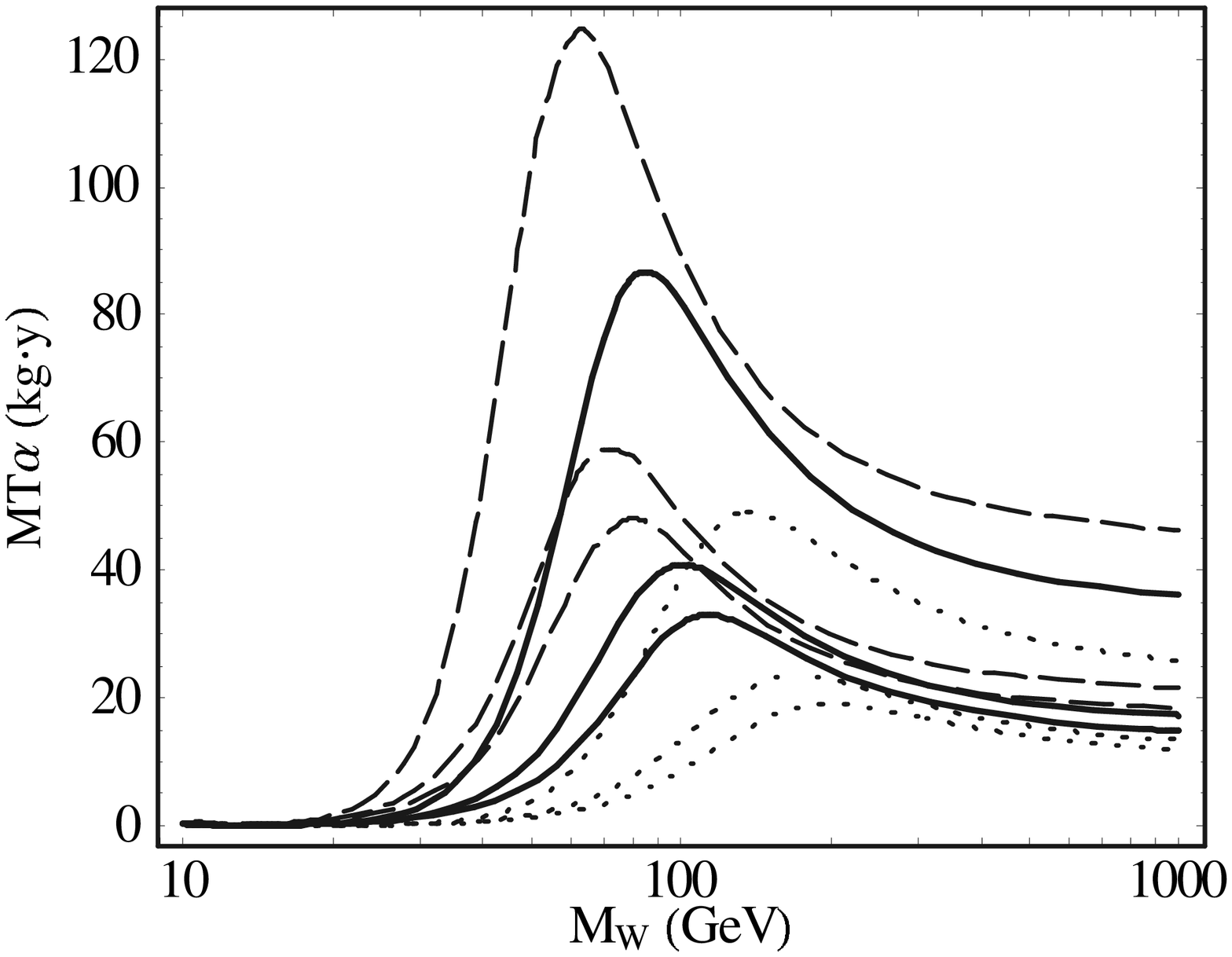,height=110mm}}
\caption{The same as in Figure 6
for a TeO$_2$
detector with threshold energy E$_{th}$=5 keV.
\label{fig:teo2_eth5_mtmin}}
\end{center}
\end{figure}

\newpage

\vspace*{5 cm}

\begin{figure}[p]
\begin{center}
\mbox{\psfig{figure=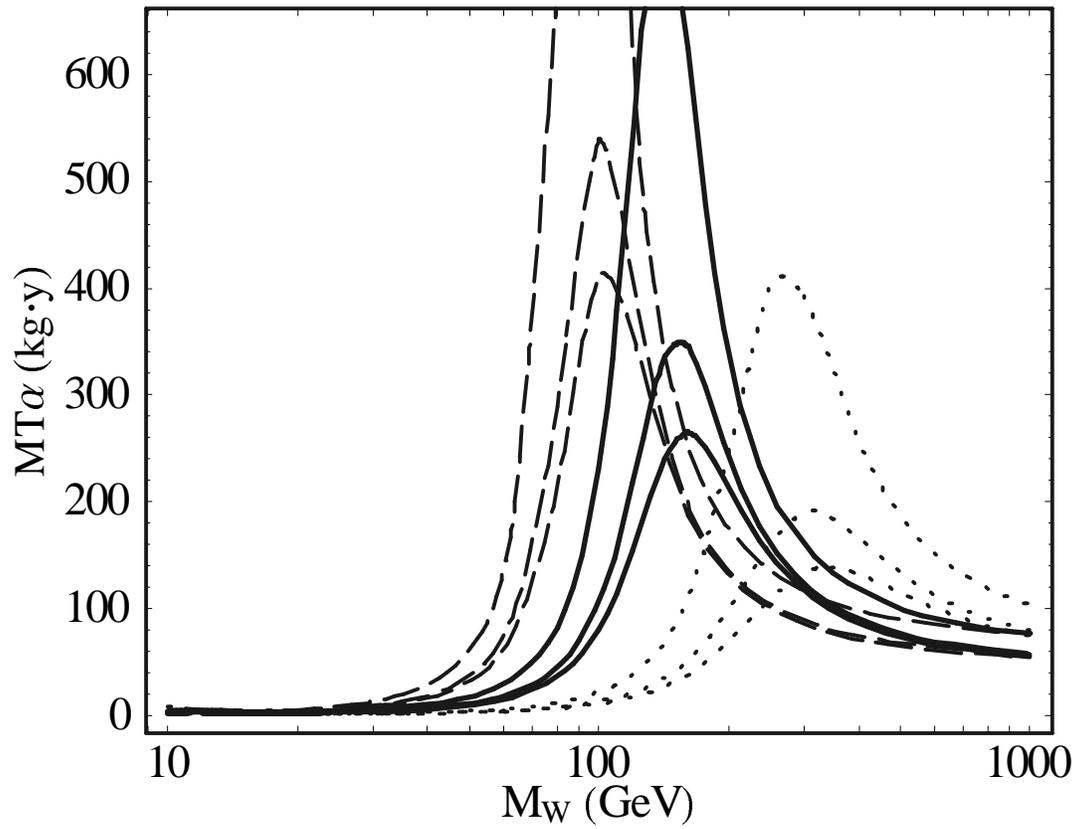,height=110mm}}
\caption{The same as in Figure 6
for an NaI
detector with threshold energy E$_{th}$=2 keV.
\label{fig:nai_eth2_mtmin}}
\end{center}
\end{figure}

\vfill\eject
\end{document}